\newcommand{\beq}{\begin{equation}}
\newcommand{\eeq}{\end{equation}}
\newcommand{\luv}{\Lambda_{UV}}
\newcommand{\as}{\alpha_s}
\newcommand{\gc}{\langle\alpha_s(G_{\mu\nu}^a)^2\rangle}
\newcommand{\lqc}{\Lambda_{QCD}}
\newcommand{\aQ}{\alpha_s(Q^2)}

\newcommand{\amu}{\alpha_s(\mu^2)}
\documentclass{ws-procs9x6}
\begin{document}

\title{Non-perturbative match of ultraviolet renormalon}

\author{V.~I. ZAKHAROV}

\address{Max-Planck Institut f\"ur Physik \\
Werner-Heisenberg Institut \\ 
F\"ohringer Ring 6, 80805, Munich\\ 
E-mail: xxz@mppmu.mpg.de}


\maketitle

\abstracts{The paper is motivated by observation of a kind of branes
in the vacuum state of the lattice SU(2) gluodynamics. The branes
represent two-dimensional vortices 
whose total area scales in physical units while the non-Abelian action 
diverges in the ultraviolet.
We consider the question whether effects of the branes
can be accommodated into the continuum theory.
We demonstrate that at least in case of the gluon condensate (plaquette
action) and of the heavy quark potential
the contribution of the branes 
corresponds to the ultraviolet renormalon. Thus, the vortices
might represent  a non-perturbative
match of the ultraviolet renormalon. Such an identification
constrains, in turn, properties of the branes.}

\section{Introduction}

Recently it has been discovered that monopoles and central vortices--
non-perturbative fluctuations commonly considered
responsible for the confinement--
have non-trivial structure in the ultraviolet.
Namely both the monopoles, see \cite{anatomy} and
references therein, and
vortices \cite{kovalenko} are associated with an excess of the
non-Abelian action which is
divergent in the ultraviolet:
\beq\label{uvd}
\langle S_{mon}\rangle~\sim~\ln 7\cdot{L\over a}~,~~~
\langle S_{vort}\rangle~\approx~0.54\cdot{A\over a^2}~~,
\end{equation} 
where $L$ is the length of the monopole trajectory, $A$ is the
area of the vortex, $a$ is the lattice spacing representing the
ultraviolet cut off. In case of monopoles,
the overall constant, which we quote as $\ln 7$ is known actually with 
rather poor accuracy but this is not crucial for our purposes.

Naively, one would expect that monopoles and vortices with action
 (\ref{uvd})
propagate only very short distances, $L\sim a$, $A\sim a^2$. However, 
both monopoles and vortices form
clusters which percolate through
the whole of the lattice volume $V_4$. Defining the corresponding densities
as:
\beq\label{definition}
L_{perc}~\equiv~4\rho_{perc}V_4~~,~~A_{vort}~\equiv~6\rho_{vort}V_4~~,
\eeq
one finds that both the monopole density (see, e.g., \cite{muller} and
references therein) and the vortex density (for review and references
see \cite{greensite}) scale in physical units and are
independent on the lattice spacing. According to the latest 
measurements:
\beq\label{ird}
\rho_{perc}~=~7.70(8)~fm^{-3}~~,~~A_{vort}~\approx~4.0(2)~fm^{-2}~~,
\eeq
see \cite{boyko} and \cite{kovalenko}, respectively.

The data (\ref{uvd}), (\ref{ird}) imply that
the standard picture of vacuum fluctuations is to be adjusted to
incorporate fine tuning, that is coexistence of the ultraviolet and
infrared scales within the same fluctuations \cite{vz,maxim}.
A comprehensive theory of the monopoles and vortices is not yet in
sight mainly because the monopoles and vortices are defined not in terms
of the original Yang-Mills fields but rather in terms of projected
fields, for review see, e.g., \cite{greensite}.

While lattice data on the monopoles and vortices have been accumulating
since long,
the discovery of the ultraviolet divergences (\ref{uvd}) makes the
challenge to the theory much more direct. Indeed, it is commonly believed
that in an asymptotically free theory 
all the ultraviolet divergences can be understood from first principles. 
And it is worth emphasizing at this point that the brane properties we
are considering, that is (\ref{uvd}), (\ref{ird}) are perfectly 
$SU(2)$ invariant. 

We will argue that, indeed, starting from the 
continuum theory it is possible to derive strong constraints on the 
ultraviolet properties of non-perturbative fluctuations
\footnote{we briefly mentioned the constraints in the talks
\cite{vz1}.}. Moreover,
 the data can be confronted
with the constraints without knowledge of  the anatomy
of the branes in terms of the original fields.   

The idea of the derivation is as follows.
The continuum theory is well defined at short distances,
while the branes are extended objects. Thus, we will project the
effect of the  branes onto matrix elements calculable in the
continuum theory. In particular, we will concentrate on the gluon
condensate and heavy quark potential at short distances.
  
Because of the asymptotic freedom the parton-model predictions
are a reliable zero-order approximation for observables of this kind. 
If one tries to derive a complete answer, then the outcome 
is an infinite perturbative series plus power-like corrections. 
The power corrections are the price
for the factorial growth of the coefficients
of perturbative expansions \cite{pth}, for
review see, e.g., \cite{review}.

We will argue that the branes with the newly discovered
properties do fit this scheme
providing a match to the so called ultraviolet renormalon. It has never
been foreseen, though, that a non-perturbative completion of the ultraviolet
renormalon could be a brane with an action density divergent in the
ultraviolet. Still, the new addition seems to fit well
the phenomenology of the power corrections. 
In turn, identification of
the branes with a non-perturbative counterpart of the ultraviolet renormalon
puts constraints on properties of the branes.

Although our main interest is the effects of the branes, we begin 
in Sect. 2 with a
rather detailed discussion of the $Q^{-2}$, or ultraviolet-renormalon
related power corrections. In particular, we argue that there are two
complementary ways to account for these corrections. 
Either one confines oneself to low-order
perturbation theory and adds the quadratic corrections ad hoc, or one is
prepared to consider high orders in perturbation theory (sometimes, say, up to ten loops)
and then the $Q^{-2}$ corrections are implicitly contained  in the
perturbative sum. While we do not suggest any new explanation of a
particular effect, our overall conclusions vary from what can be found
in the literature. The point is that commonly it is assumed,
explicitly or tacitly, that the two procedures outlined above are
mutually inconsistent. 

In Sect. 3 we review phenomenological evidence on the $Q^{-2}$
corrections and argue that the data are consistent with the
theoretical conclusions.
 
In Sect. 4 we  argue that the branes provide a non-perturbative match
to the ultraviolet renormalon. In view of the discussion above this
means,
in turn, that the branes--when projected onto local matrix elements--
are dual to high orders of perturbation theory. In Sect. 5 we discuss
also the corresponding 
constraints on properties of the  vortices. 

\section{Perturbative expansions}

\subsection{Divergences of perturbative expansions}

Standard analysis of divergences of perturbative expansions
\footnote{This subsection is a very sketchy review.
For details, see, e.g., \cite{review}.}
introduces hierarchy of three scales,

$$\lqc^2~\ll~Q^2~\ll~\luv^2~,$$
where $\lqc$ and  $\luv$ are the standard infrared and ultraviolet scales
while the intermediate scale $Q$
is specific for a particular process considered.
 Then a generic perturbative expansion for
an observable $\langle~ O~\rangle$ looks as:
\beq \label{pert}
\langle~ O~\rangle~= ~(parton~model)\cdot
\big(1~+~\sum_{n=1}^{\infty}a_n\alpha_s^n(Q^2)~\big)~~,
\end{equation}
where for simplicity of presentation we normalized
the anomalous dimension of the operator $O$ to zero.
Note also that $\aQ\ll 1$.

In fact, expansions (\ref{pert}) are only formal
since the coefficients $a_n$ grow factorially at large $n$:
\beq\label{growth}
|a_n|~\sim~c_i^n\cdot n!~~,
\end{equation}
where $c_i$ are constants. Moreover, there are a few sources
of the growth (\ref{growth}) and, respectively, $c_i$
can take on various values.
First, the growth of the coefficients $a_n$ may reflect the
growth of the number of perturbative graphs in high orders,
which is combinatorial in nature. The value of $c_i$ is determined then
by the classical action of an instanton solution.
On the other hand, the divergence (\ref{growth}) can be triggered by
a single graph of $n$-th order. This is a renormalon-type graph 
which contains $n$ insertions of vacuum polarization into a gauge boson
propagator.

The factorial growth of $a_n$ implies that the expansion (\ref{pert})
is asymptotic at best. Which means, in turn, 
that (\ref{pert}) cannot approximate
a physical quantity to accuracy better than
\beq\label{uncert}
\Delta~\sim~\exp\big(-1/c_i\aQ\big)~\sim~
\Big({\lqc^2\over Q^2}\Big)^{b_0/c_i}~~,
\end{equation}
where $b_0$ is the first coefficient in the $\beta$-function,
$$Q^2{d\over dQ^2}\aQ~=~-b_0\alpha_s^2(Q^2)~-~b_1\alpha_s^3(Q^2)~-...~~~~,$$ 
and we accounted for the fact that $\aQ$ is logarithmically small at
large $Q^2$. 

To compensate for these intrinsic uncertainties one modifies
the original expansion (\ref{pert}) by adding power corrections
with unknown coefficients:
\beq\label{pert1}
\langle~ O~\rangle~= ~(parton~model)\cdot
\Big(1~+~\sum_{n=1}^{\infty}\tilde{a}_n\alpha_s^n(Q^2)~~+~\sum_{k}
b_k(\lqc/Q)^k~~\Big),
\end{equation}
where powers $k$ are determined in terms of $c_k$ entering Eq. (\ref{growth})
and the factorial divergences are removed from the modified 
perturbative coefficients $\tilde{a}_n$, for further discussion
see, e.g., \cite{mueller}.

\subsection{Borel summation}

So far, we discussed only behavior of absolute values of the 
coefficients $a_n$. Now, if
there is sign oscillation, 
\beq\label{sign}
a_n\sim (-1)^nn!c_i^n~~,
\end{equation}
then the sum is Borel summable 
while if $a_n\sim (+1)^nn!c_i^n~$ there is no way
at all to define the sum. 
In case of an asymptotically free theory the sign oscillation
(\ref{sign})
is characteristic for the ultraviolet renormalons.
For the first ultraviolet renormalon,
\beq\label{uvrenormalon}
(a_n)_{uv}~\sim~n!b_0^n(-1)^n~~,
\end{equation}
where $b_0$ is the first coefficient in the $\beta$-function.
Usually one does not reserve for any uncertainty in case of a
Borel summable series.

The criterion of Borel summability might look too formal. 
Let us mention, therefore, that there exist
also intuitive reasons to believe that there is no intrinsic
uncertainty of perturbative series due to the ultraviolet renormalons.
Indeed, one can readily check that the physical meaning of the
power terms is that they correspond to contributions of non-typical virtual
momenta $k^2_{virt}$
which are either small or large compared to $Q^2$. 
The ultraviolet renormalons  are associated with
$$k^2_{virt}~\gg~Q^2~~,$$
while infrared renormalons correspond to 
$$k^2_{virt}~\ll~Q^2~~.$$
Furthermore, it is rather obvious that the contribution of large momenta in
asymptotically free theories is calculable exactly. Then, there could
be no intrinsic uncertainty due to the ultraviolet renormalon.

On a more technical level, the argument runs as follows \cite{beneke1}.
If one changes normalization of the coupling used in the expansion
(\ref{pert}) from $g^2(Q^2)$ to $g^2(\mu^2)$ then the uncertainty
(\ref{uncert}) associated with the ultraviolet renormalon
changes to:
\beq
\Delta (\mu^2)~\sim~{\lqc^2\over \mu^2}\cdot{Q^2\over \mu^2}~~.
\end{equation}
Thus using a normalization point $\mu^2\gg Q^2$ one can get rid of the
corresponding
power-like uncertainty. There is no similar trick in case of the
infrared renormalons, signaling that the corresponding power terms
is a genuine uncertainty of the perturbative expansion.

\subsection{Non-perturbative match to infrared renormalon}

In phenomenological applications, we  will concentrate mostly on two 
particular examples, that is, gluon condensate $\gc$,
where $G^a_{\mu\nu}$ is the non-Abelian field-strength tensor,
and the heavy quark potential $V(r)$ at short distances, $r\to 0$.
Keeping only the leading
power
corrections we have \cite{svz}:
\beq\label{gc}
\langle~ (G_{\mu\nu}^a)^2~\rangle~\approx~(N^2_c-1)
\cdot a^{-4}\big(1~+~\sum
a_n\as^n(a^2)~+~c_4(\lqc\cdot a)^4~\big)~,
\end{equation}
where $N_c$ is the number of colors and $a$ is the lattice spacing, 
and \cite{balitsky,ligeti,akhoury}:
\beq\label{potential}
\lim_{r\to 0}V(r)~\approx~-{(N_c^2-1)\over 2N_c} {\alpha_s(r)\over   
r}\big(1~+~\sum_na_n\alpha_s^n(r)~+~
c_3 \lqc^3r^3~\big)~.
\end{equation} 
A salient feature of the predictions (\ref{gc}) and (\ref{potential}) is
the
absence of quadratic corrections of order $(\lqc\cdot a)^2$, 
$(\lqc\cdot r)^2$.

It is worth emphasizing that although the power corrections are so to say
detected through pure perturbative graphs the actual vacuum fluctuations
which dominate the power corrections in (\ref{gc}), 
(\ref{potential}) can well be genuinely non-perturbative. 
In particular, the leading power corrections both in (\ref{gc}) and
(\ref{potential}) correspond to the so called infrared renormalon,
a perturbative graph with many iterations of a loop insertion
into the gluon propagator. 
However, the same dependence on $\lqc$ is provided by instantons:
\beq
\langle\alpha_s(G_{\mu\nu}^a)^2\rangle_{inst}~\approx
~(const)\lqc^4~.
\end{equation}
Moreover, there are well developed and, in many respect, 
successful models of vacuum
which assume the instanton dominance, see \cite{shuryak} and references
therein. 

It is important to realize that the instanton dominance in no way 
contradicts (\ref{gc}), (\ref{potential}). Indeed, 
analysis of the $n$-dependence of
the coefficients $a_n$ fixes the position of a singularity in the
Borel plane and predicts in this way that the leading power correction
is proportional to $\lqc^4$. This prediction is confirmed
by evaluating the instanton contribution. Instantons might
enhance the correction numerically and make the whole analysis more
tractable.

\subsection{Status of the ultraviolet renormalon}

Turning back to the ultraviolet renormalon, -- which is most relevant to
the present paper, -- one might feel that the summary 
presented above is in fact self-contradictory. Indeed, if we start with
expansion in $\aQ$, see (\ref{pert}) then we would reserve for  a
$(\lqc^2/Q^2)$ correction corresponding to the ultraviolet
renormalon. Which would be the leading power correction.
Even if we apply summation a la Borel, a $(\lqc^2/Q^2)$ contribution
would
arise as a result of the summation of the divergent tail of the
perturbative
series. On the other hand, if we start with expanding in $\amu$,
$\mu^2 \gg Q^2$, there is no intrinsic uncertainty to the perturbative
series which would correspond to the ultraviolet renormalon. 

The contradiction is superficial, however.
Namely, if one uses expansion in $\amu$, then the $(\lqc^2/Q^2)$ terms  are
to emerge as a result of explicit summation of high orders of
perturbation theory. Indeed, in terms of the original expansion
(\ref{pert})
the ultraviolet renormalon corresponds to terms of order 
$$N_{uv}~\approx~{1\over b_0\aQ}~~.$$
In terms of the $\amu$ expansion one needs to keep even higher
orders.

It is worth emphasizing that an explicit calculation of the
 ultraviolet
renormalon in gluodynamics is not straightforward at all
\cite{arkady}. The point is that the ultraviolet-renormalon  divergence
(\ref{uvrenormalon}) is in fact related not only to the simplest chain graph 
but to a whole class
of graphs,
and no explicit calculation is possible.

To summarize, one expects that the quadratic corrections are present
but they are hidden either in the postulated procedure of summation a la Borel or 
in explicit summation of higher orders. It is worth emphasizing that 
numerical enhancement of the $(\lqc^2/Q^2)$ terms is not ruled out by
this analysis. However, such an enhancement is not suggested either.

\section{Status of $Q^{-2}$ corrections}

\subsection{Power corrections from the infrared}

There exists huge literature on power corrections, for review see, e.g.,
\cite{gc}. Here, we will mention only a few points  relevant to our
discussion.

The most standard sum rules \cite{svz} apply to the two-point functions
\beq\label{correlator}
\Pi_j(Q^2)~=~i\int\exp (iqx)\langle 0\vert T\{j(x),j(0)\}\vert 0\rangle,~~
(q^2~\equiv~-Q^2)~~,
\end{equation}
where $j(x)$ is a current constructed on the quark and gluon fields and
for simplicity we omitted the Lorenz indices. Moreover, one usually
considers the Borel transform of (\ref{correlator}),
$$
\Pi_j(M^2)~=~[Q^{2n}/(n-1)!](-d/dQ^2)^n \Pi_j(Q^2)~,$$
where $n\to\infty,Q^2\to \infty$ with $Q^2/n\equiv M^2$ fixed.

Somewhat symbolically, the sum rules then read:
\beq\label{sumrule}
\Pi_j(M^2)\approx({\rm parton~ model})\Big(1~+~{a_j\over\\ln (M^2/\lqc^2)}~+~
{c_j\over M^4}~+~O\big((\ln (M^2/\lqc^2))^{-2},M^{-6}\big)\Big)~~,
\end{equation}
where the constants $a_j,c_j$ depend on the channel, i.e. on the current $j$.
Terms of order $1/\ln M^2$ and $M^{-4}$ are the first perturbative correction
and contribution of the gluon condensate, respectively. Note that we
kept
only first order correction in $\alpha(M^2)$.
We could have included, say, one more perturbative term,
but the sum-rule approach is still based on the assumption that
the effect of the power corrections is most important numerically.

The physical meaning of the power corrections in (\ref{sumrule}) is
that they parameterize contribution of large distances of order
$\lqc^{-1}$ in terms of local operators, like $(G_{\mu\nu}^a)^2$. 
The technical means behind is the operator product expansion (OPE)
and there the procedure is well defined and
unambiguous \cite{nsvz2}.

While there is no doubt that (\ref{sumrule}) retains the leading power
corrections if $M^2$ is large enough, it was noticed quite early \cite{nsvz3}
that in some channels there could exist terms which start as sub-leading power corrections   
at large $M^2$ but numerically dominate at moderate $M^2\sim
few~GeV^2$. Moreover, realistically such corrections  are 
associated with so called direct instantons \cite{shuryak}, for a very 
recent and
impressive example see \cite{degrand}.

\subsection{The case for large  $Q^{-2}$ corrections}

While sum rules (\ref{sumrule}) claim many successes, 
there has also been accumulating evidence suggesting
introduction of $Q^{-2}$ corrections which apparently go beyond the standard 
picture summarized in the preceding subsection.

To begin with, the instanton-dominated model of vacuum does not 
reproduce the confining potential for 
heavy external quarks \cite{shuryak2}. On the other hand, lattice studies
indicate that the so called Cornell potential for heavy quarks, 
\beq\label{cornell}
V_{Q\bar{Q}}(r)~\approx~-{c_V\over r}~+~\sigma\cdot r~~,
\end{equation}
describes the data at all the distances $r$, for a review see
\cite{bali1}. At large distances, the linear
potential provides confinement. For our purposes here it is 
crucial, however, that the fit (\ref{cornell}
suggests also that if we start with short distances
the leading power correction 
to the Coulomb-like interaction at `moderate' distances $r$ 
is of order $\sigma\cdot r^2$.
 This observation is one of the motivations to
modify the r.h.s. of Eq. (\ref{potential}) by
adding a $(\lqc\cdot r)^2$ term (or $\delta V(r)\sim \lqc^2 r$)
\cite{akhoury,shevchenko}.

The gluon condensate, also, contains large quadratic corrections
if the perturbative series is truncated at first terms. We will
discuss this example in detail in the next subsection. 

Another dramatic example of large quadratic corrections is found in
Ref. \cite{shuryak3} and concerns instanton density as function
of the instanton size $\rho$.
Namely, according to the standard operator product expansion,
underlying also the sum rules (\ref{sumrule}), one has \cite{svz1}:
\beq\label{instantons}
dn(\rho)~\approx~dn_{pert}(\rho)
\Big(1+{\pi^4\rho^4\over 2g^4}\langle 0\vert 
(G^a_{\mu\nu})^2\vert 0\rangle~\Big)~~,
\end{equation}
where $g(\rho)$ is the gauge coupling
and the salient feature of (\ref{instantons}) is, again, absence of
a quadratic correction. However, it is demonstrated in \cite{shuryak3}
that the data on the instanton density, to the contrary, 
unequivocally require
a quadratic term. The data on the instanton density are described by
an effective action:
\beq\label{instanton}
S_{eff}~\approx~{8\pi^2\over g^2(\rho^2)}~+~c_{\rho}\sigma \rho^2~~,~~
c_{\rho}~\approx~2\pi~,
\eeq
where the first term is the standard action in the quasiclassical
 approximation and  $\sigma$ is
the string tension. The expression (\ref{instanton}) describes the variation of the instanton
density $dn(\rho)$
by about four orders of magnitude.

There exist further examples when quadratic corrections
improve phenomenological fits, see, e.g., \cite{narison2}.
However,  the $Q^{-2}$ corrections are associated with short
distances and there is no 
substitution for the OPE which would allow to relate quadratic
corrections to various quantities in a model-independent way.
One of the ideas put forward in different contents, see, e.g.,
\cite{shirkov} is a universal change of the running coupling
by a $Q^{-2}$ term. In particular, one argues sometimes \cite{shirkov}
that analyticity requires
removal of the Landau pole from the running coupling. The use of the
modified,
or `analytical' coupling,
\beq\label{analytical}
\aQ_{anal}~\approx~{1\over b_0}\Big({1\over \ln
(Q^2/\lqc^2)}~+~{\lqc^2\over \lqc^2-Q^2}\Big)~~,
\eeq 
then introduces a  $1/Q^2$ correction at large $Q^2$.
An apparent weak point is that imposing analyticity
on higher orders in $\as$
introduces further $Q^{-2}$ corrections which are not suppressed 
\cite{vz5}.

One can try to generalize (\ref{analytical})
and think in terms of a universal coupling which includes
a $Q^{-2}$ term. Let us note that the quadratic corrections 
in (\ref{cornell}) and
(\ref{instanton}) differ numerically by a factor of 1.5 and agree
in sign if the correction is ascribed to a universal
coupling. This can be considered as a success
of the model. Anyhow, till now there was no more convincing 
theoretical estimates of the quadratic term in (\ref{instanton}).
 
Quantitatively, the model with a short-distance gluon mass 
\cite{chetyrkin} turns
to be most successful. According to the model, the sum rules
(\ref{sumrule}) are modified as
\beq\label{sumrule1}
\Pi_j(M^2)~\approx~({\rm parton~ model})\Big(1~+~{a_j\over\ln M^2/\lqc}~+~
{b_j\over M^2}~+~
{c_j\over M^4}~+...~\Big)~~,
\end{equation}
where the coefficients $b_j$ are proportional to the gluon mass squared
and are calculable for any channel. The model cleared all the hurdles 
known. For example, one finds \cite{narison2}:
\beq
b_{\pi}~\approx~4b_{\rho}~~,
\end{equation}
and this resolves a long standing puzzle \cite{nsvz3} of the analysis
of the sum rules in the $\pi$-meson channel. 

To summarize,  it seems reasonable, -- as far as phenomenology
of the confinement-related effects is concerned, -- to replace  
the standard free gluon propagator by a propagator which reproduces the whole
of the potential (\ref{cornell}) already in the zero-order
approximation of perturbation theory. On the theoretical side,
such a  program 
is a refinement of the variational approach \cite{thorn} and is
outlined in \cite{coupling}. Practically, full higher order calculations
within such a scheme are known to be  very cumbersome \cite{konishi}.
As far as the power corrections are concerned, this approach
amounts to introducing a tachyonic gluon mass at short distances
already in the lowest order of perturbation theory.
Which is easy to implement and successful phenomenologically \cite{chetyrkin}.

\subsection{Perturbative - non-perturbative `ambiguity'}

Note that all the phenomenological successes mentioned in the previous
section refer  to the fits with only first order perturbative
contribution retained, along with a $Q^{-2}$ term.
The successes claimed refer to moderate $Q^2$ where the power
correction becomes sizable.
However, one could argue that if we start with large $Q^2$
then to extract a small power-like
corrections one
needs to subtract as many orders of perturbation theory as possible.
Then, the question is how this would affect the fits to the $Q^{-2}$
correction at small distances.

Numerically, this problem was studied in greatest detail  in case of
the gluon condensate on the lattice \cite{burgio,rakow}. 
In terms of the lattice formulation the gluon condensate is nothing
else but the average plaquette action \cite{digiacomo}.  
The result of the calculations can be summarized in the following way.
Represent the plaquette action $\langle P\rangle $ as:
\beq\label{plaquette}
\langle P\rangle ~\approx~P_{pert}^N~+~b^Na^2\lqc^2~+c^Na^4\lqc^4~~,
\end{equation}
where the average plaquette action $\langle P\rangle $
is measurable directly on the lattice and is known to high accuracy,
$P_{pert}^N$ is the perturbative contribution calculated 
up to order N:
\beq\label{pn}
P_{pert}^N~\equiv~1~-~\sum_{n=1}^{n=N}p_ng^{2n}~~,\eeq
and, finally coefficients $b^N,c^N$ are fitting parameters
whose value depends on the number of loops $N$. Moreover, the form of
the fitting function (\ref{plaquette})
is rather suggested by the data 
than imposed because of theoretical considerations.

The conclusion is that up to ten loops, $N=10$ it is the quadratic correction
which is seen on the plots while $c^N$ are consistent with zero.
However, the value of $b^N$ decreases monotonically with growing $N$ 
\cite{burgio}
and somewhere
after 10 loops becomes consistent with zero while the $\lqc^4$ term
finally shows up \cite{rakow}.  Moreover, the
emerging value of the $\lqc^4$ term is consistent with current
phenomenological estimates from the sum rules \cite{rakow}.

Turn now to another example of a large $Q^{-2}$ correction, that is 
the heavy quark potential, see (\ref{cornell}). The first attempt to 
analyze the effect of higher perturbative corrections on the linear
term at short distances was undertaken in Ref. \cite{bali2}. Defining
the potential as
\beq\label{attempt}
V_{Q\bar{Q}}(q)~\equiv~-C_F{4\pi \alpha_V(q)\over q^2}~~,
\eeq
and using first terms in the perturbative expansion of $\alpha_V$
which
are known explicitly,
\beq
\alpha_V(q)~=~\alpha_{\bar{MS}}(q)\big(1~+~a_1\alpha_{\bar{MS}}(q)~+~
a_2\alpha^2_{\bar{MS}}(q)~\big)~~,
\eeq
one finds that the linear piece in the potential at short distances is
5 times larger than it would follow from (\ref{cornell}). In other
words, the result depends crucially on the subtraction procedure.
To proceed further, one is to invoke  a model.

In particular, it was suggested \cite{pineda}
to saturate higher orders by the leading infrared renormalon
evaluated  in the large-$b_0$ approximation:
\beq\label{pineda}
\alpha_V(q)~\approx~\alpha_{\bar{MS}}(q)\big(1~+~a_1\alpha_{\bar{MS}}~+~
a_2\alpha^2_{\bar{MS}}
~+~\sum_{n=3}^N a_n^{ren}\alpha^n_{\bar{MS}}\big)
\eeq
where $N\sim 3/2b_0\alpha_{\bar{MS}}$. 
The potential observed on the lattice is reproduced then
by (\ref{pineda})
including $\delta V(r)\sim \sigma r$ at short distances. Moreover,
the result is stable against reasonable variations in $N$.
Note that no explicit
$Q^{-2}$ terms are to be introduced within this procedure
since high orders of perturbation theory are presumably
accounted for explicitly.

\subsection{Complementary ways of describing the  $Q^{-2}$ terms}

Thus, the lesson from calculations of
the perturbative gluon condensate 
is that one can either
approximate total matrix elements by a low-order 
perturbative contribution plus a quadratic correction,
or by high-order perturbative contributions plus a quartic 
correction. 

Our central point, which motivated us to review the evidence in favor
and against non-standard $Q^{-2}$ corrections, is that there is no 
contradiction between the two approaches. Moreover, existence of the
dual
descriptions of the $Q^{-2}$ terms is expected in fact on pure
theoretical grounds, see discussion of the status of the ultraviolet
renormalon in Sect. 2.4 . 

Theoretically, it is known 
that no ad hoc $Q^{-2}$ terms are allowed
to be added to untruncated perturbative series \cite{beneke1}. 
However, then one
is to be prepared to calculate high orders indeed. 
In the only case when such a calculation turns possible
(that is, the gluon condensate \cite{burgio,rakow}) theoretical expectations 
are fully confirmed. Namely, lowest orders plus a quadratic correction
give reasonable fits
\cite{burgio,rakow}. On the other hand, if many loops are accounted
for, then there is no need for an ad hoc power correction any longer
\cite{rakow}. What remains unanswered at this point is
why the $Q^{-2}$ terms, -- described in
one or the other way,-- are important 
phenomenologically. As we will argue later, observation of the branes
seems to provide us with a key to answer this question.

\section{Branes and power corrections}

\subsection{Quadratic correction to the gluon condensate}

Monopoles and vortices are detected, for a given configuration of the
vacuum fields, for the whole of the lattice. Thus, they are seen 
as a nonlocal structure. Moreover, they are manifestly
non-perturbative. Indeed, 
the  probability $\theta(plaq)$ for a particular plaquette to belong to a brane
has been found to be proportional to:
\beq\label{area1}
\theta(plaq)~\approx~(const)\exp(~-1/b_0g^2(a))~\sim ~(a\cdot \lqc)^2~.
\end{equation}
On the other hand, the branes have an  ultraviolet divergent tension
which assumes a kind of locality.

 To make contact with the continuum theory it is useful to evaluate
contribution of the branes into local or quasi-local matrix elements.
The gluon condensate (\ref{gc}) turns to be the easiest case. Indeed,
combining Eqs (\ref{ird}) and (\ref{uvd}) one gets for the
contribution of the vortices to the gluon condensate:
\beq\langle~(G_{\mu\nu}^a)^2~\rangle_{vort}~\approx~0.3~GeV^2~a^{-2}~~,
\end{equation}
which matches the ultraviolet renormalon.
The beauty of this result is that all the quantities considered are
manifestly
gauge invariant.

\subsection{Quadratic correction to the heavy quark potential}

 In case of the heavy quark potential, one can also argue that
the branes match the ultraviolet renormalon, or the $Q^{-2}$ correction.
Indeed, it is well known that both monopoles and vortices generate
the linear potential for heavy quarks.
In particular, in case of the monopoles \cite{shiba}:
\beq\label{piece}
\big(~V_{Q\bar{Q}}(r)\big)_{mon}~\approx~ \sigma_{mon}\cdot r~~,
\end{equation}
where $\sigma$ is close numerically to the string tension in
the full potential (\ref{cornell}). Moreover, the potential
(\ref{piece})
is linear at all distances tested, beginning with $r=a$. 
The reservation is that the observation (\ref{piece}) refers to the
Abelian projected potential. However, the approximation is known
to be valid for the quarks in
the fundamental representation (
for detailed discussion see, e.g., \cite{muller}).

\subsection{Probing properties of the branes}

To produce a linear potential (\ref{piece}) at short distances, 
vacuum fluctuations are to satisfy highly non-trivial constraints.
First, the size of fluctuations is to be small \cite{akhoury,cgpz}.
Indeed consider again for simplicity the Abelian 
projection. Then
the potential
can be represented as an integral over the overlap of electric fields 
associated with the external charges charges $\pm Q_{el}$:
\beq
V(r)~=~{1\over 4\pi}\int {\bf E}({\bf r'})\cdot {\bf E}({\bf r}+{\bf
r'})d^3r'~~,
\eeq
where ${\bf r}$ is the radius-vector connecting the charges. Imagine
that there is  a  change in the field at distances of order $R$.
In the applications, $R\sim (\lqc)^{-1}$. At distances
$R\gg r$ the field of the charges is a dipole field: 
$$|{\bf E}_{dip}|~\sim~Qr/R^3~. $$  Respectively, the change in the 
potential is of the order:
\beq
\delta V(r)~\sim~Q^2 r^2 \int_R^{\infty}d^3r'/(r')^6~\sim~{Q^2
r^2\over R^3}~,
\eeq 
and we would conclude that the non-perturbative
potential  at small distances is proportional to $r^2$, 
see also Eq. (\ref{potential}).

Therefore, the standard estimate might not work only in case that 
fluctuations responsible  for the confinement have size comparable or
less than the distance between the quarks, $r$. Monopoles and
vortices with the action (\ref{uvd}) do satisfy this constraint
since the ultraviolet divergence in the action implies size
of order $a$.

To appreciate another constraint on the properties of the vortices,
consider the $Z_2$ projection,
(for a review see \cite{greensite}). Then the area law for the Wilson loop is
derived from the assumption that the probabilities for plaquettes in
the plane of the Wilson loop to belong to the vortex are uncorrelated
\cite{casimir}. Indeed, in this case the Wilson loop on average is given by:
\beq\label{random}
\langle W(C)\rangle ~=~\Pi_{plaq}\big((1-f)+f(-1)\big)\langle
W_0(C)\rangle~~,
\eeq
where the product is taken over all plaquettes belonging to the 
minimal area $A$ spanned on the contour $C$,
$f$ is the probability for a plaquette to belong to the vortex,
and  $\langle W_0(C)\rangle$ is the value of the loop with the
constraint that no vortices pierce the minimal area.
Then,
\beq
\langle W(C)\rangle ~\approx~\exp(- \sigma A/a^2)\langle
W_0(C)\rangle~~,
\eeq
where the string tension $\sigma~=~-\ln (1-2f)$ and $A$ is the
minimal area in the lattice units. Note that 
$f\sim (a\lqc)^2$, see Eq. (\ref{area1}).

Now, we are discussing the case when the contour $C$ has one of
dimensions very small, $A\sim T\cdot r$ where $T$ is the time
extension and $r\to 0$. Eq. (\ref{random}) applies only if
the vortex remains random even at distances comparable to the lattice
spacing $a$. Which implies, in turn, that the entropy factor for the
vortex grows exponentially with its area:
\beq\label{entropy}
(Entropy)_{vort}~\sim ~\exp(~c A/a^2)~~.
\eeq
This is indeed true for the branes observed on the lattice, see
\cite{kovalenko}.

So far, we have not discussed corrections to the instanton density,
see (\ref{instanton}) since no quantitative framework is known
in this case. Let us only note that the physics behind the standard
prediction (\ref{instantons}) is similar to the case of the heavy
quark potential. Indeed, in case of instantons we deal with color
dipoles in d=4. Again, the OPE prediction (\ref{instantons}) follows
from the assumption that the confining fields are soft and  modify
the original instanton field at distances of order $(\lqc)^{-1}$. 
The validity of (\ref{instanton}) implies, therefore, that there exist
small-size vacuum fluctuations. There are no other 
(non-perturbative) candidates but the branes.

\subsection{Branes and perturbative series}

The original ultraviolet renormalon sequence of the perturbative
coefficients is given by (\ref{uvrenormalon}). The $n!$ factor arises
then from integrals over virtual momenta of the type:
\beq\label{largemomenta}
\int_{\sim Q^2}^{\luv^2}{dk^2 (\ln k^2/Q^2)^n\over k^4}~~.
\eeq
Such a contribution is related to a single graph with
$n$ insertions of the vacuum polarization into a gauge boson
propagator. However, this is not a single source of 
contributions of order (\ref{uvrenormalon}) \cite{arkady}. Namely,
one can reserve, say, for one loop insertion less 
because then one gains in terms of
number
of graphs. In this way there arise contributions of order
$$a_n~\sim~(n-1)!\cdot l~~, ~~l~\sim~ n~,$$
where the first factor is due the mechanism (\ref{largemomenta})
and the factor $l\sim n$ is combinatorial.
Thus, one  gradually switches to evaluating both the number of graphs and
powers of the logs. This two-parameter problem becomes practically
intractable after a few steps. 

It is worth emphasizing that in case of the power corrections to
the gluon condensate there is no mechanism (\ref{largemomenta})
at all. Indeed the expansion is in terms of $g^2(\luv^2)$
and there are no virtual momenta larger than the normalization point.

Thus, the only mechanism left for having a large quadratic correction
is a large number of graphs. Moreover, large power-like terms cannot
be associated now with sign oscillation. Indeed, sign oscillation plus
expansion in $g^2(\luv^2)$ would result in a negligible
contribution. Thus, we come to conclusion that the series
associated with the quadratic correction should not oscillate in sign. 
Only then we can expect that the series corresponds to a sizable
quadratic correction.

Let us check the expectations against the only known 
example of an explicit calculation of a `long' perturbative
series, that is, turn to the case of the gluon condensate
in the lattice $SU(3)$ gluodynamics.
According to \cite{rakow}:
\beq\label{rakow}
r_n~\equiv~{p_n\over p_{n-1}}~\approx~u\Big(1~-~{1+q\over n+s}\Big)~~,
\eeq
where the coefficients $p_n$ are introduced in
(\ref{pn}) the numerical values of the parameters are:
$$
u~=~0.961(9),~~q~=~0.99(7)~~,~~s~=~0.44(10)~~.$$
The quadratic correction is affected by  $n\eqsim 10$.

We see, indeed, that the perturbative series has no sign oscillation.

So far, two regular mechanisms for generating same-sign perturbative
expansions were discussed \cite{review}. Namely, infrared renormalons
and  perturbative-vacuum instability due to instantons. 
In both cases the ratio $r_n$ grows at large $n$ linearly
with $n$. In case of the leading infrared renormalon:
\beq\label{rren} 
r_n^{ren}~\approx~{nb_0\over 8\pi}~~.
\eeq
Remarkably enough, in the crucial region of $n\sim 10$ this
ratio is still  smaller than (\ref{rakow}) and the infrared
renormalon
catches up with (\ref{rakow}) only around $n\sim 25$ \cite{rakow}. 
In case of instanton-related divergence 
the ratio $r_n$ is even smaller than in case (\ref{rren}):
\beq\label{rninst}
r_n^{inst}~\approx~{2nb_0\over 44\pi} ~~.
\eeq

From the theoretical point of view, both the instantons and infrared
renormalons are irrelevant to the $Q^{-2}$ corrections. Thus,
it is gratifying that the perturbative series (\ref{rakow}) looks
indeed
different. The series (\ref{rakow}) is convergent for $|g^2|~<~u$
and seems to exhibit a novel mechanism of generating same-sign
perturbative series.  

\subsection{Crossover}

The non-analyticity in $g^2$ indicated by (\ref{rakow})
might be a reflection of the crossover transition \cite{rakow}. 
Indeed, the irregular
behavior of the specific heat associated with the crossover is
known since long \cite{lautrup}. Note also that at the crossover
the branes, with the properties known in the weak-coupling region,
disappear. Indeed, in the strong-coupling region, that is beyond
the crossover there is no scale $\lqc$ at all and the basic properties
of the branes, like (\ref{area1}) make no sense.

Thus, there arises the following tentative picture of the 
mechanism behind the $Q^{-2}$ corrections. Because
of the crossover transition, there appear corrections to the standard running
of the coupling at mass scale of about $GeV$. This change 
in the running is manifested
phenomenologically in various ways. Thus, there
is evidence for a `freezing' of the effective coupling,
for review see, e.g., \cite{badalian}. In particular, the
following effective coupling:
\beq
\alpha_{eff}(r)~\approx~{1\over b_0l_{mod}}\Big(1~+~{b_1\over b_0^2}{ln
l_{mod}\over l_{mod}}\Big)^{-1}~~,
\eeq
where
$$l_{mod}~\approx~\ln {{1+r^2\cdot (GeV)^2\over r^2\cdot\lqc^2}}~,$$
fits well various pieces of data \cite{badalian}.

On the non-perturbative side, 
the branes with properties (\ref{uvd}), (\ref{ird})
emerge around the crossover
(if one moves towards weak coupling). The simplest
and qualitative fit is therefore lowest order in perturbation theory
plus effect of the branes. This fit is justified in the region
close to the crossover -- and mostly the lattice measurements are
performed for $\beta$ not far from the crossover.
On the other hand, the same effect is to be described by perturbation
theory since the measurements are already in the weak-coupling domain
and there is no singularity which would block the perturbative 
expansion from being valid in the region close to the crossover
(approached from the weak-coupling side). 
The price is that, the closer is the crossover, the more
terms in the perturbative expansion are to be kept.
Evaluating so many perturbative terms is practically impossible
except for the case of the gluon condensate \cite{rakow}.

The perturbative series corresponding
to the $Q^{-2}$ correction exhibits sign coherence, similar to the case
of a classical solution. The series is not divergent in large orders,
however.
That is, branes are no classical solution. Numerically,
the perturbative expansion coefficients 
in the crucial region of  
$n$ are even larger than for known divergent series (infrared
renormalon, instantons). This is correlated with  the fact that the
branes apparently represent strongly aligned vacuum fluctuations.

\section{Constraints on the branes}

\subsection{Consistency of the branes with asymptotic freedom}

In the preceding section we could convince ourselves that
to match the ultraviolet renormalon the branes are to have
highly non-trivial properties like (\ref{uvd}), (\ref{area1}),
(\ref{entropy}). All these properties, crucial to fit the ultraviolet
renormalon, reveal a point-like facet of the branes.
Now we are reversing the question and ask whether this point-likeness
is consistent with the asymptotic freedom. In particular, 
the action for the lattice monopoles is proportional to $a^{-1}$, the
same  as
for point-like particles.  At first sight, it is unavoidable
that accepting (\ref{uvd}) in the limit
$a\to 0$ is equivalent to introducing new particles at short
distances. Appearance of such particles would be inconsistent with the
asymptotic freedom. 

To have a closer look at the problem it is useful to translate
the data on the monopole trajectories into a conventional
field theoretic language. To describe monopoles one then introduces
a magnetically charged field $\phi_M$. Moreover, since 
the monopoles are observed as trajectories, it is natural to use the
polymer representation of field theory, see, in particular, \cite{sym}.
Proceeding in this way one can derive 
(see \cite{vz,maxim} and references therein):
\beq\label{ve}
\langle 0| ~|\phi_M|^2~|0\rangle~=~{a\over 8}\big(\rho_{perc}~+~
\rho_{fin})~~,
\eeq
where density of the percolating cluster, 
$\rho_{perc}$ is defined in (\ref{definition}) while $\rho_{fin}$
is related in a similar way to the length of the finite 
monopole clusters.

The total density of the monopole clusters
 can be measured directly
\cite{muller,boyko}. It is crucial that the total monopole
density diverges as $a^{-1}$ at small $ a$:
\beq\label{aminus}
\rho_{perc}~+~\rho_{fin}~\approx~(const)\lqc^3~+~(const^{`}){\lqc^2\over
a}~~.
\eeq
Finally, we get:
\beq\label{lqc2}
\langle 0| ~|\phi_M|^2~|0\rangle~\approx~(const^{`})\lqc^2/8~
\approx~0.8~(fm)^{-2}.
\eeq
Thus, $\langle 0| ~|\phi_M|^2~|0\rangle$ contains no ultraviolet
divergence and is, therefore, perfectly consistent with the
asymptotic freedom which does not allow to add new particles. 

To appreciate the geometrical meaning of the observation (\ref{aminus})
turn to the simplest case of uncorrelated percolation. One
introduces then a probability $p$ for a link to be ``open'', that is to
belong to a monopole trajectory in our notations. Then at a critical
value $p_{cr}$ there arises an infinite, or percolating cluster.
The density of this percolating cluster is vanishing at the point
of the phase transition to the percolation. In the supercritical
phase where $p>p_{cr}$ this density
\beq\label{percolating}
\rho_{perc}~\sim~(p-p_{cr})^{\alpha}~,
\eeq
where the critical exponent $\alpha >0$. 

While the density (\ref{percolating}) is vanishing at $p=p_{cr}$,
no non-analyticity can happen in the total density. At the point
of the phase transition, it is given simply by:
\beq\label{totall}
\rho_{tot}~=~{p_{cr}\over a^{d-1}}~,
\eeq
where $d$ is the number of dimensions. 

Substituting (\ref{totall}) into (\ref{ve}) we recover in case
of $d=4$ the standard quadratic divergence,
$$\langle 0| ~|\phi_M|^2~|0\rangle\sim a^{-2}$$
which is so familiar from the case of Higgs particles.
However, the data (\ref{aminus}) indicate only $a^{-1}$ behavior of
the total monopole density. Which geometrically means that the
monopoles spread over a $d=2$ subspace of the total $d=2$ space.
(Of course, the $d=2$ subspace is not a plane but rather a surface
percolating itself through the $d=4$ space).

Thus, we come to an amusing conclusion that it is the existence of the
branes which eliminates a potential quadratic divergence in (\ref{ve}).
Note that the fact that the monopoles are associated with
the vortices (whose total area scales) was observed first
in Ref. \cite{giedt}. Later the phenomenon was confirmed
for various values of $a$ \cite{kovalenko,syritsyn}. In view of the
ultraviolet divergence in the monopole action, see (\ref{uvd}),
this association of the monopoles with vortices becomes
absolutely crucial for the consistency with the asymptotic
freedom.

Our final  comment concerning (\ref{lqc2}) is that the 
vacuum expectation value (\ref{lqc2})
is perfectly gauge invariant. Gauge invariant condensates of 
dimension two were widely discussed recently, see, e.g. 
\cite{kondo}. The beauty of the relation (\ref{lqc2}) is that
it does not contain ultraviolet divergences which plague 
local condensates formulated in terms of the gluon fields.  
Thus, the vacuum expectation value (\ref{lqc2}) could be 
related directly to the non-perturbative
part \cite{boucaud} of the gluonic dimension two condensates.
No explicit relation of this kind is known, however.

\subsection{Matching of the branes with ultraviolet renormalon}

The lattice data on the branes, see, in particular (\ref{uvd}), (\ref{ird})
are compelling but pure numerical. Since structure of the branes in
terms of the
original gluon fields is not known, it is very difficult, for example, to
explain (\ref{area1}) theoretically. Matching of the branes 
with the ultraviolet renormalon brings  
constraints on the branes
which turn to be quite restrictive. 
In particular, to match the ultraviolet
renormalon
in case of the gluon condensate the branes are to obviously satisfy the
following constraint:
\beq\label{constraint}
\theta_{plaq}\cdot {({\rm thickness})\over a}
\cdot\big(({\rm plaquette
~action})\cdot a^4\big)~\sim~\lqc^2a^{2}~~,
\eeq
where $\theta_{plaq}$ is the probability for a given plaquette to
belong
to the vortex (see (\ref{area1})) and the  thickness is understood
in terms of the distribution of the excess of the $SU(2)$ action.
The presently available data indicate \cite{kovalenko}
that $\theta_{plaq}\sim (\lqc\cdot a)^2$,
the plaquette action is of order $a^{-4}$ and the
thickness is equal to the resolution, that is $a$.
(All the quantities are on the average.)  
Such a regime is fully consistent with (\ref{constraint}) and can,
therefore, persist in the limit $a\to 0$ as well.

There is another and subtler point. The non-Abelian action,
albeit ultraviolet divergent,
is known to be much smaller than the projected one. In case
of monopoles this was emphasized, in particular,  in
\cite{suganuma}:
\beq\label{suganuma}
S^{mon}_{non-Ab}~\sim ~{L\over a}~~\ll~~
S^{mon}_{Ab}~\sim~~{1\over g^2}{L\over a}~,
\eeq
where $S^{mon}_{Ab}$ is the action in the Abelian projection, 
i.e. the same as for the Dirac monopole. Similar inequality
holds in case of the vortices. 

Now, we can derive (\ref{suganuma}). Indeed, the
branes are dual to high orders of perturbation theory. 
Which means that the action associated with the branes is
is not allowed to be parametrically  more singular than the perturbative
fluctuations. This rules out the monopole action of order
$S_{mon}\sim g^{-2}\cdot L/a$, see Eq. (\ref{suganuma}). In other words,
branes cannot be removed from the vacuum  without affecting 
perturbative fields as well.

\subsection{Casimir scaling~~}

The lattice data on the quark potential (for review see, e.g.,  \cite{bali1})
reveal a strikingly simple picture. Lack of a same simple theoretical
explanation looks as a puzzle. In fact there exist rather a few
puzzles to be explained:

a) at relatively short distances the potential is well approximated by
a Coulomb like piece plus a linear potential;

b) the linear piece continues, with the same slope to large
distances;

c) Casimir scaling. 

The phenomenon of the Casimir scaling \cite{casimir} is that 
at intermediate distances static 
potentials for quarks  belonging to various representations $D$ 
of the color group
are proportional to each other:
\beq 
V_D(r)~\approx~{C_D\over C_F}V_F(r)~~,
\eeq
where $C_D$ is the eigenvalue of the quadratic Casimir operator
$$C_D~=~TrT^D_aT^D_a$$
of the representation, $C_F$ is the value of $C_D$ for the fundamental
representation and $V_F(r)$ is the static quark potential in case of
heavy quarks in the fundamental representation.

Explanation of the observed Casimir scaling is a challenge to
theory, for review and models see \cite{casimir,casimir1}. 
It is not obvious either that
observation of the branes brings a transparent solution to the problem.
We still feel that the consideration
of the branes does introduce new elements into understanding of the
Casimir scaling.

As far as the observation b) above is concerned, a new point is that the
branes look the same random at all scales tested. Namely, the
properties (\ref{uvd}), (\ref{ird}), (\ref{entropy}) reiterate
themselves for all the lattice spacings. Since there is no 
scale involved, except for $\lqc$,
see Eq. (\ref{area1}) the same slope $\sigma\sim \lqc^2$
governs the brane-induced potential at all the distances. An explicit 
realization
of this idea is provided by Eq. (\ref{random}). 

Thus, the branes allow to shift the emphasis in explaining 
the Casimir scaling 
from large to  short 
distances. Concerning short distances,the linear potential
is calculable, in principle, perturbatively.
In the two-loop approximation
(that is,  three terms in the expansion of $V(r)$ in $\as$)
the Casimir scaling is checked by explicit calculations \cite{peter}:
\beq\label{twoloop}
V_D^{two~loop}(r)~=~{C_D\over C_F}V_F^{two-loop}(r)~~.
\eeq
However, the linear term at short distances is sensitive to even
higher orders (for a discussion see
Sect 3.3) and one has to turn to models. In particular,
the model with a tachyonic gluon mass \cite{chetyrkin} immediately
gives:
\beq
\delta V_D(r)~=~ {C_D\alpha_s\over 2}|m^2_g|r~~,
\eeq
reproducing the Casimir scaling.

More generally, the branes are to be dual to the ultraviolet
renormalons. In turn, the factorization properties
of the ultraviolet renormalon, see \cite{beneke2} and references therein,
result in the Casimir scaling.
However, within such a framework it is difficult to expect that the Casimir
scaling would hold to high accuracy.

To summarize, consideration of branes allows to reduce the problem
of the Casimir scaling to the problem of evaluating the linear
potential at short distances. The latter problem is perturbative 
in nature and the first two loop corrections are known to exhibit
the Casimir scaling. Higher orders can be estimated only within models.

\section{Conclusions}

We have demonstrated that the monopoles and vortices are responsible
for the ultraviolet-renormalon type corrections at short distances.
In turn, the $Q^{-2}$ corrections are dual to high orders of
perturbation theory.
Phenomenologically, such corrections are known to be welcome. 
Combination of the ultraviolet and infrared factors typical for the
ultraviolet renormalon appears to be a reflection of the fine tuning
for the monopoles and vortices as non-local objects. Vortices and
monopoles  appear, 
in this context, as non-perturbative counterpart of the ultraviolet renormalon.
Highly non-trivial constraints on the properties of the branes implied
by this identification are satisfied by the lattice data.

The overall conclusion is that the fine tuning observed
on presently available lattices, see in particular (\ref{uvd}),
(\ref{ird}), could be the true asymptotic in the limit $a\to 0$.

In a broader context, fluctuations appearing as topological and
suppressed in one formulation of a theory can become fundamental
entities in a dual formulation of the same theory. If it is true
in case of  $SU(2)$ Yang-Mills theory, then
the hint is that the branes could appear in the dual formulation.
Where by branes we understand (see above) $d=2$ surfaces populated
by the monopoles (tachyonic mode) and living on a $d=4 $ Euclidean
space. Generically, this observation agrees with recent 
and well known proposals \cite{maldacena} on theories
dual to YM theories. To the best of our knowledge,
no direct comparison of the two approaches is possible, however.

\section*{Acknowledgements}

This paper grew out of discussions of the results of the lattice
measurements presented in 
\cite{anatomy,boyko,kovalenko,syritsyn}. I am thankful to V.G. Bornyakov,
P.Yu. Boyko, M.N. Chernodub, F.V. Gubarev, A.V. Kovalenko, M.I. Polikarpov, 
T. Suzuki and S.N. Syritsyn for collaboration and thorough discussions.

I am thankful to A. DiGiacomo, J. Greensite, 
Yu.A. Simonov, A.A. Slavnov, L. Stodolsky, E.T. Tomboulis,
and K. Van Acoleyen for interesting discussions. 

The paper was completed while the author was visiting the Yukawa
Institute for Theoretical Physics of the Kyoto University.
I am thankful to Prof. T. Kugo and other members of the group for
the hospitality.

\end{document}